\pdfoutput=1
\documentclass[iop,apjl,tighten]{emulateapj}
\usepackage{graphicx,enumitem}
\usepackage[bottom]{footmisc}
\newcommand{\tabincell}[2]{\begin{tabular}{@{}#1@{}}#2\end{tabular}}
\usepackage{threeparttable}
\usepackage{multirow}

\usepackage{xcolor}
\usepackage{hyperref}
\hypersetup{
   colorlinks,
   linkcolor={blue!88!black!80},
   citecolor={blue!88!black!80},
   urlcolor={blue!88!black!80}
}

\newcommand{\lya}{Ly$\alpha$}
\newcommand{\HeII}{He\,\textsc{ii}}
\newcommand{\OII}{[O\,\textsc{ii}]}
\newcommand{\Hb}{H$\beta$}
\newcommand{\OIII}{[O\,\textsc{iii}]}

\newcommand{\CIV}{C\,\textsc{iv}}
\newcommand{\NV}{N\,\textsc{v}}

\shorttitle{z $\sim$ 7.0 LAEs}
\shortauthors{Hu W. et al.}

\begin{document}

\title{First Spectroscopic Confirmations  of z $\sim$ 7.0 Ly$\alpha$ Emitting Galaxies  in the LAGER Survey}

\email{urverda@mail.ustc.edu.cn, jxw@ustc.edu.cn, zhengzy@shao.ac.cn}

\author{
  Weida Hu\altaffilmark{1, 5}, Junxian Wang\altaffilmark{1, 5}, Zhen-Ya Zheng\altaffilmark{2,3,6}, Sangeeta Malhotra\altaffilmark{4,7}, Leopoldo Infante\altaffilmark{3,6}, James Rhoads\altaffilmark{4,7}, Alicia Gonzalez\altaffilmark{4}, Alistair R. Walker\altaffilmark{8}, Linhua Jiang\altaffilmark{9}, Chunyan Jiang\altaffilmark{2}, 
  Pascale Hibon\altaffilmark{10},  L. Felipe Barrientos\altaffilmark{3,6}, Steven Finkelstein\altaffilmark{11}, Gaspar Galaz\altaffilmark{3,6}, Wenyong Kang \altaffilmark{1, 5}, Xu Kong\altaffilmark{1,5}, Vithal Tilvi\altaffilmark{4}, Huan Yang\altaffilmark{1,4}, XianZhong Zheng\altaffilmark{12}
}
\affil{
$^1$CAS Key Laboratory for Research in Galaxies and Cosmology, Department of Astronomy, University of Science and Technology of China,
Hefei, Anhui 230026, China; urverda@mail.ustc.edu.cn, jxw@ustc.edu.cn\\
$^2$CAS Key Laboratory for Research in Galaxies and Cosmology, Shanghai Astronomical Observatory, Shanghai 200030, China; zhengzy@shao.ac.cn\\
%$^4$Institute of Astrophysics and Center for Astroengineering, Pontificia Universidad Catolica de Chile, 7820436 Santiago, Chile; linfante@astro.puc.cl \\
$^3$Centro de Astroingenier\'ia, Facultad de F\'isica, Pontificia Universidad Cat\'olica de Chile, Santiago, Chile\\
$^4$School of Earth and Space Exploration, Arizona State University, Tempe, AZ 85287, USA; Sangeeta.Malhotra@asu.edu, James.Rhoads@asu.edu\\
$^5$School of Astronomy and Space Science, University of Science and Technology of China, Hefei 230026, China\\
$^6$Instituto de Astrof\'isica,  Facultad de F\'isica, Pontificia Universidad Cat\'olica de Chile, Santiago, Chile\\
$^7$Astrophysics Science Division, Goddard Space Flight Center, 8800 Greenbelt Road, Greenbelt, 
Maryland 20771, USA;\\
%$^5$Chinese Academy of Sciences South America Center for Astronomy, 7591245 Santiago, Chile \\
$^8$Cerro Tololo Inter-American Observatory, Casilla 603, La Serena, Chile\\
$^9$The Kavli Institute for Astronomy and Astrophysics, Peking University, Beijing, 100871, China\\
%$^9$N\'ucleo de Astronom\'ia, Facultad de Ingenier\'ia y Ciencias , Universidad Diego Portales, Av. Ej\'ercito 441, Santiago, Chile\\
$^{10}$European Southern Observatory, Alonso de Cordova 3107, Casilla 19001, Santiago, Chile\\
$^{11}$Department of Astronomy, The University of Texas at Austin, Austin, TX 78712, USA\\
$^{12}$Purple Mountain Observatory, Chinese Academy of Sciences, Nanjing 210008, China\\
}

\begin{abstract}

Narrowband imaging is a highly successful approach for finding large numbers of high redshift Ly$\alpha$ emitting galaxies (LAEs) up to $z\sim6.6$. However, at $z\gtrsim7$ there are as yet only 3 narrowband selected LAEs with spectroscopic confirmations (two at $z\sim6.9-7.0$, one at $z\sim7.3$), which hinders extensive studies on cosmic reionization and galaxy evolution at this key epoch. 
We have selected 23 candidate $z\sim 6.9$ LAEs in COSMOS field with the large area narrowband survey  LAGER (Lyman-Alpha Galaxies at the End of Reionization). In this work we present spectroscopic followup observations of 12 candidates using IMACS on Magellan.   For 9 of these, the observations are sufficiently deep to detect the expected lines.
Ly$\alpha$ emission lines are identified in six sources (yielding a success rate of 2/3),  including 3 luminous LAEs with Ly$\alpha$ luminosities of
$L_{\mathrm{Ly\alpha}} \sim$ 10$^{43.5}$ erg s$^{-1}$, {the highest among known spectroscopically confirmed galaxies at  $\gtrsim7.0$}. This triples the sample size of spectroscopically confirmed narrowband selected LAEs at $z\gtrsim7$, and confirms the bright end bump in the Ly$\alpha$ luminosity function we previously derived based on the photometric sample, supporting a patchy reionization scenario. 
Two luminous LAEs appear physically linked with projected distance of 1.1 pMpc and velocity difference of $\sim$ 170 km s$^{-1}$. They likely sit in a common ionized bubble produced by themselves or with close neighbors, which reduces the IGM attenuation of Ly$\alpha$.  
A tentative narrow \NV$\lambda$1240 line is seen in one source, hinting at activity of a central massive black hole with metal rich line emitting gas.

\end{abstract}


\section{Introduction}

In the past two decades, narrowband imaging surveys have been proven a highly efficient approach for finding high redshift Ly$\alpha$ emitting galaxies (LAEs).
Many LAEs from $z \sim 2$ to $\sim$ 6.6 have been selected via narrowband imaging and spectroscopically identified \citep[e.g.][]{Hu1996, Hu2004, Hu2010, Wang2009, Rhoads2003, Rhoads2004, Dawson2004, Dawson2007, Ouchi2005, Ouchi2008, Ouchi2010, Tapken2006, Westra2006, Shibuya2014, Shibuya2017, Zheng2016}.

As Ly$\alpha$ photons can be resonantly scattered by the neutral hydrogen in the IGM in the early universe, Ly$\alpha$ emitters provide a powerful probe of cosmic reionization.
Comparison of Ly$\alpha$ luminosity functions at $z \sim 5.7$ and 6.5 \citep{Malhotra2004, Hu2010, Ouchi2010, Kashikawa2011,Matthee2015}
suggest a mostly ionized IGM at $z\sim6.5$ ( x$_{HI}$ $<$ 0.3).
However, at $z\gtrsim 7$, when the dominant phase of cosmic reionization took place, only a few candidate LAEs have been 
selected from various narrowband surveys \citep[at redshifts of $\sim$ 6.9 -- 7.0, 7.3, 7.7 and 8.8,][]{Iye2006, Ota2008, Ota2012a, Ota2012b, Hibon2010, Hibon2011, Hibon2012, Shibuya2012, Konno2014, Tilvi2010, Krug2012, Matthee2014}.
More importantly, only 3 of them have been spectroscopically confirmed \citep[two at $z \sim 6.9$, one at $z\sim 7.3$, ][]{Iye2006, Rhoads2012, Shibuya2012}, demonstrating the significant challenge in the search 
for LAEs at $z\gtrsim 7$. 
The challenge is at least partly due to IGM attenuation of Ly$\alpha$ line flux at $z\gtrsim 7$ \citep[e.g.][]{Zheng2017, Konno2014, Shibuya2012, Tilvi2010, Matthee2014}.
Much larger samples of  $z\gtrsim7$ LAEs, particularly spectroscopically confirmed ones, are essential to probe the physics of reionization and galaxy formation/evolution in the early universe. 

LAGER (Lyman-Alpha Galaxies in the Epoch of Reionization) is an ongoing large narrowband survey for LAEs at $z \sim 6.9$,
using the Dark Energy Camera (DECam, with FOV $\sim$ 3 deg$^2$) on the NOAO-CTIO 4m Blanco telescope
with an optimally designed custom narrowband filter NB964 (FWHM $\sim$ 90\AA).

In the first LAGER field (COSMOS), we have selected 23 LAE candidates at $z \sim 6.9$ \citep{Zheng2017}.
The Ly$\alpha$ luminosity function shows a rapid faint-end evolution from $z\sim6.9$ to $z\sim6.6$ and 5.7, suggesting an IGM neutral hydrogen fraction of x$_{HI}$=0.4--0.6 at $z\sim6.9$.
More strikingly, the LF shows a clear excess in the bright end, including 4 of the brightest candidates \citep[with L$_{Ly\alpha}$ $\sim$ 10$^{43.4}$ erg s$^{-1}$,][]{Zheng2017}.
Such bright end LF excess suggests that reionization could be patchy, and those 4 luminous candidates located in ionized bubbles which reduce the IGM attenuation of Ly$\alpha$ lines. 

In this paper we present spectroscopic followup observations of 12 of the candidate $z\sim6.9$ LAEs. 
In \S2 we present the observations and data reduction. The spectra and identifications are given in \S3, followed by discussion 
in \S4. 
{Throughout this work, we adopt a flat $\Lambda$CDM cosmology with $\Omega_{\mathrm{m}}=0.3$, $\Omega_{\mathrm{\Lambda}}=0.7$ and H$_0=70$ km s$^{-1}$ Mpc$^{-1}$.}

\section{Observations and Data Reduction}

\subsection{Spectroscopic Observations}
With the narrowband filter NB964 ($\lambda_c = 9642$\AA, FWHM $\sim$ 90 \AA) and the DECam mounted on the CTIO Blanco 4m telescope, we have obtained 34 hrs NB exposure in COSMOS field, reaching a 3$\sigma$ limiting AB magnitude of 25.6 (2\arcsec\ diameter aperture).
Candidate $z\sim6.9$ LAEs were selected with flux excess between NB964 and deep z band images (z - NB964 $\ge$ 1.0). Foreground contaminations are efficiently excluded using deep bluer broadband images. The final sample includes 23 candidates \citep[for details see][]{Zheng2017}.
{All of our candidates have $L_{\mathrm{Ly\alpha}} \geq$ $4.5\times10^{42}$ erg s$^{-1}$ and rest-frame EW$_{0,\mathrm{Ly\alpha}}$ $\geq$ 10 \AA.}

We carried out spectroscopic follow-up for 12 of 23 candidate LAEs using the Inamori Magellan Areal Camera and Spectrograph (IMACS) on the 6.5m Magellan I Baade Telescope on 2017 February 6-8 and March 21-22. 
We used the IMACS $f/2$ camera (with FOV of 27\arcmin\ diameter) with 300-line red-blazed grism. IMACS masks with 1\arcsec\ slit width were designed to cover candidate LAEs, foreground fillers, 
6 late-type standard stars 
from UltraVISTA catalog \citep{Muzzin2013}
for on-mask flux calibration, and at least 11 alignment stars. 
Two masks were observed during the run, with each mask covering 6 candidate LAEs (LAE-4,6,8,13,17,18 on Mask 1, and  LAE-1,2,3,11,21,23 on Mask 2). 
We ultimately obtained 390 minutes IMACS exposure for Mask 1 and 340 minutes exposure for Mask 2 (Table \ref{tbl:sso}). 
The seeing was $0\farcs5-0\farcs8$ during the February run and $0\farcs8-1\farcs2$ in the March run,
however the February run (Mask 2) was affected by significant moon light pollution (with the moon 81\%--95\% illuminated and 65$^\circ$--38$^\circ$ from the field on 2017 Feb 6--8).

\begin{table*}[]
\caption{Summary of IMACS Spectroscopic Observations}
\label{tbl:sso}
\centering
\begin{threeparttable}
\begin{tabular}{c l l c l l}
\hline
\hline
Mask ID & \tabincell{c}{Observation Data} & \tabincell{c}{$t$* (s)} & \tabincell{c}{Number of Exposures} & \tabincell{c}{$t_{total}$** (s)} & \tabincell{c}{Seeing}\\
\hline
Mask 1 & 2017 Mar 21 & 600 & 1 & 600 & $1\farcs2$\\
Mask 1 & 2017 Mar 21-22 & 1200 & 19 & 22800 & $0\farcs8-1\farcs2$\\
Mask 2 & 2017 Feb 6-8 & 1200 & 17 & 20400 & $0\farcs5-0\farcs8$\\
\hline
\end{tabular}
\begin{tablenotes}
\item[*] Exposure time per frame.
\item[**] Total exposure time.
\end{tablenotes}
\end{threeparttable}
\end{table*}

\subsection{Data Reduction and Analysis}
We used COSMOS2\footnote{http://code.obs.carnegiescience.edu/cosmos} software package for IMACS data reduction. COSMOS2 does bias subtraction, flat fielding using quartz lamp exposures, wavelength solution and calibration using arc lamp exposures, and sky subtraction.  Finally, it extracts a two-dimensional spectrum for each slit. After standard data reduction steps, we remove cosmic rays using L.A.Cosmic \citep{vanDokkum2001}, and stacked spectra from individual frames using a weighted-stack method to maximize the S$/$N of the co-added spectra. The signal-to-noise ratio of an exposure is inversely proportional to the RMS background variation.  Thus, it is reasonable to built a wavelength-dependent weighting map for an exposure based on this background variation as  measured from source-free pixels along the slit direction at each wavelength:
\begin{equation}
w_i(\lambda) = \frac {1}{\sigma_i^2},
\end{equation}
where $w_i(\lambda)$ is the weight at $\lambda$  in the $i$th exposure,  and $\sigma_i$ is the RMS background variation at $\lambda$. Finally, we obtained a two-dimensional stacked spectrum for each slit. We then extracted one-dimensional spectra from stacked two-dimensional spectra using IRAF task APALL with an extraction window of $1\farcs2$ along the spectral traces. 
We flux-calibrated the spectrum of each candidate LAE using the spectra of the six stars included on the slit mask, using the stars' known magnitudes and colors to estimate their intrinsic spectral energy distributions.

\begin{table*}[]
\caption{Spectroscopical Properties of the Six LAEs}
\label{tbl:ssp}
\centering
\begin{tabular}{c c c c c c c c c c c c}
\hline 
\hline
Name  & RA & DEC &  \tabincell{c}{Redshift$^{a}$} & \tabincell{c}{$f_{\mathrm{Ly}\alpha}^b$ } & \tabincell{c}{$f_{\mathrm{Ly}\alpha}^c$   } &  EW$^c_\mathrm{{0,phot}}$ & EW$^d_\mathrm{{0,spec}}$ 
& \tabincell{c}{Skewness${^e}$}
& \tabincell{c}{M$_{1500}^f$} & \tabincell{c}{FWHM${^g_{\mathrm{int}}}$} & S/N \\
\hline
LAE-1 & 10:02:06.0 & +2:06:46.1 & 6.936 & $3.86\pm0.27$ &7.79 & $115\pm14$ & $57\pm6$ & $7.0\pm2.0$ & $-21.3\pm0.2$ & $251^{+30}_{-31}$ & 14.3 \\ %97
LAE-2  & 10:03:10.5 & +2:12:30.8 & 6.922 & $2.36\pm0.29$ & 5.86&  $>40$ & $>16$ & $-6.0\pm3.5$ & $>-23.6$ & $134^{+48}_{-65}$ & 8.1 \\ %128
LAE-3 & 10:01:53.5 & +2:04:59.6 &6.931 & $2.89\pm0.28$ & 5.11 &  $73\pm9$ & $41\pm6$ & $6.1\pm3.00$ & $-21.2\pm0.2$ &$364^{+45}_{-46}$ & 10.3 \\ %92
LAE-14 & 9:58:45.2 & +2:31:29.2 & 6.924 & $2.10\pm0.24$ & 1.48 &  $>134$ & $>190$ & $0.8\pm1.8$ & $>-20.4$ & $195^{+33}_{-36}$ & 8.6\\ %19
LAE-17 & 9:59:21.7 & +2:14:53.2 & 6.885 & $1.26\pm0.15$ & 1.20 &  $24\pm6$ & $25\pm4 $ & $0.2\pm1.9$ & $-21.7\pm0.1$ & $<155$ & 8.4 \\ %37 
LAE-18 & 9:59:59.8 & +2:29:06.5 & 6.925 & $0.85\pm0.19$ & 1.21 &  $>36$ & $>25$ & $1.0\pm4.7$ & $>-21.5$ & $<173$ & 4.5\\%56 
\hline
\end{tabular}
\begin{tablenotes}
\item[$^a$] $^a$ The redshift is determined from the Ly$\alpha$ emission line fitted with Gaussian profile;
\item[$^b$] $^b$ Spectroscopically measured Ly$\alpha$ line flux ($\mathrm{10^{-17}\ erg\ s^{-1}\ cm^{-2} }$); 
\item[$^c$] $^c$ Photometrically measured Ly$\alpha$ line flux and rest frame EW$_0$ using NB964 and Ultra-VISTA Y band (DR3) photometry \citep{Zheng2017, McCracken2012}, in unit of $\mathrm{10^{-17}\ erg\ s^{-1}\ cm^{-2} }$ and \AA; Lower limits to EW$_0$ are obtained using 2$\sigma$ upper limits to Y band magnitudes;
\item[$^d$] $^d$ {Rest frame }Ly$\alpha$ line EW$_0$ derived using spectroscopical line flux and Y band based continuum flux (\AA).
\item[$^e$] $^e$ {Weighted skewness, as defined in \citet{Kashikawa2006}.}
\item[$^f$] $^f$ {UV magnitude, measured using Ultra-VISTA J band (DR3) photometry \citep{McCracken2012}.}
\item[$^g$] $^g$ Line width corrected for the instrumental broadening, $\mathrm{km\ s^{-1}}$.
\end{tablenotes}
\end{table*}
\section{Spectroscopic Results}

\begin{figure*}
\begin{center}
\includegraphics[width=160mm]{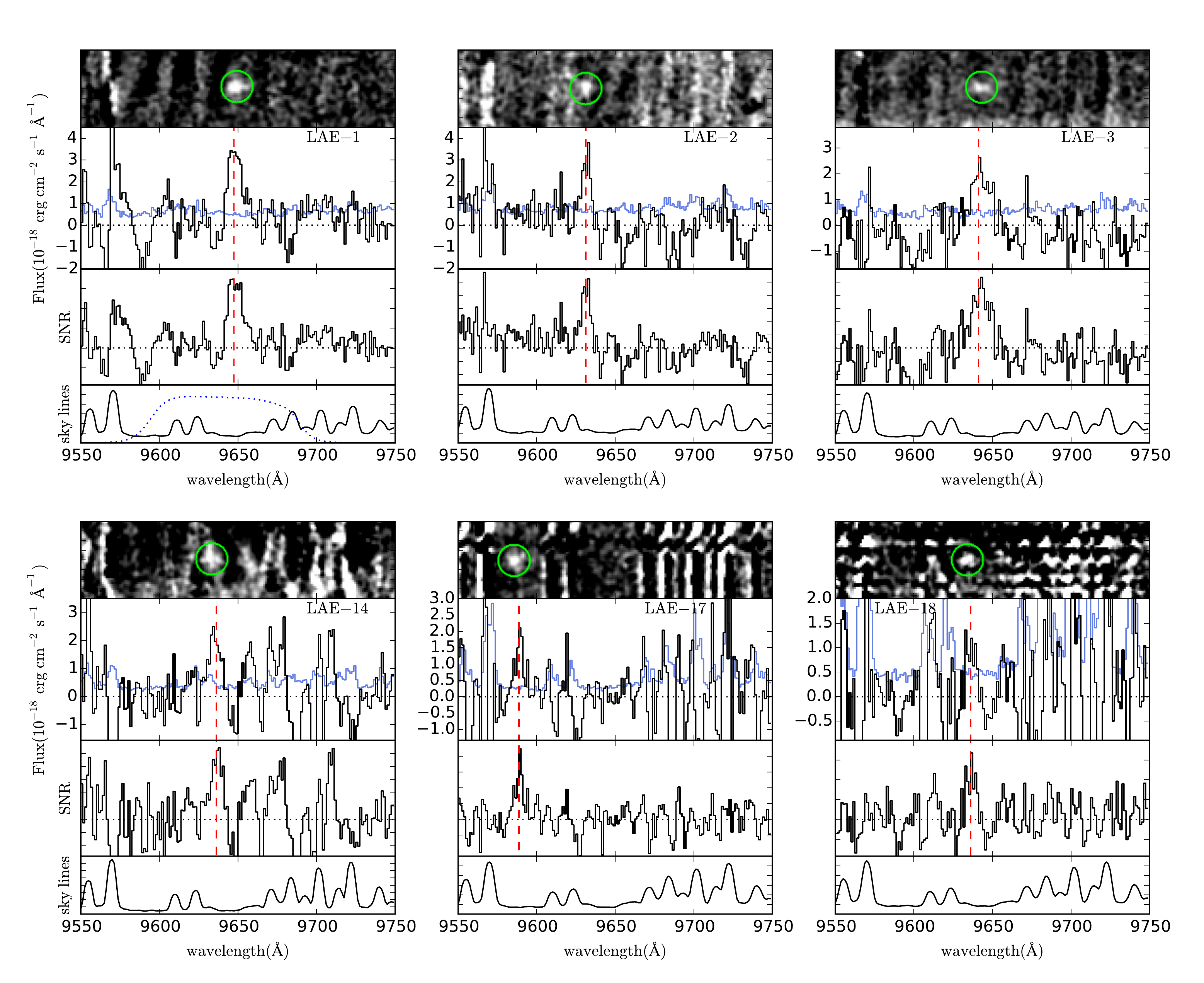}
\caption{\label{spec}Two-dimensional spectra {(smoothed with the instrumental resolution)},  one-dimensional flux and S/N spectra of the six spectroscopic confirmed Ly$\alpha$ lines. In the upper left panel, the transmission curve of NB964 is over-plotted (blue dashed line) with sky line spectrum.
The blue solid lines in each panel plot the noise spectra, both in this figure and subsequent ones.}
\end{center}
%\label{spec}
\end{figure*}

Among the 12 candidates, emission lines at expected wavelengths are detected in six sources. 
{The lines are visible in both halves of the dataset stacked separately, showing that they are not artificial features.}
We do not detect continuum from any of them. We present the 2-d and 1-d spectra of the 6 sources in Fig. \ref{spec}.
Except for LAE-18 (in which we identified two lines), these lines are the only ones we detected in the spectra.
We fitted a Gaussian profile to every detected line, and measured line fluxes using the Pyspeckit Python module \citep{Ginsburg2011}.
{The uncertainties in the line fluxes are estimated through Monte-Carlo simulations. We added random noise based on the variation along the slit direction to the 2D spectra, re-extracted the 1D spectra,  and re-measured the line fluxes. We run the simulations 10000 times and fitted a Gaussian to the flux distribution to derive the uncertainty of the flux measurement. The same technique was applied to obtain the uncertainties of other line parameters.}
We also performed weighted skewness  \citep{Kashikawa2006} measurements for those emission lines.  Weighted skewness $>$ 3 is considered a robust indication that a line is asymmetric \citep{Kashikawa2006}. Table \ref{tbl:ssp} summarizes the spectroscopic properties of our candidate LAEs.

The significant emission lines in LAE-1 and LAE-3 are clearly asymmetric with red wings, typical of high redshift Ly$\alpha$ emission lines. 
The 4 other lines are narrower and statistically consistent with symmetry,
{though the non-detections of line asymmetry in these sources might be due to low line S/Ns or an adjacent sky line (in LAE-2). }
The intrinsic line widths of the 6 sources span $\sim$ 100 -- 300 km s$^{-1}$,
{while the instrumental spectral resolution is  $\sim$ 184 km s$^{-1}$.} 
{Although line profiles alone can only robustly confirm LAE-1 and LAE-3 as LAEs, }
we identify all the six lines as Ly$\alpha$ at $z \sim$ 6.9, and low-$z$ foreground emission lines can be ruled out as follows.

\OIII $\lambda$4959 can be ruled out because the stronger \OIII $\lambda$5007 lines are not seen.
If the detected lines are \OIII $\lambda$5007,  the lower limits to the \OIII $\lambda$5007/\Hb\ ratio of most sources would be implausibly large ($>$ 10).
For LAE-17, 
if the observed line was \OIII $\lambda$5007, the lower limit of \OIII $\lambda$5007/\OII $\lambda$3727 ($>$ 11.45) would require a rather low metallicity of 12+log(O/H) $<$ 7 \citep{Nagao2006}.
However LAE-17 has a red color (z-J $>$ 1) inconsistent with a young and low metallicity galaxy at $z \sim 0.9$.

\Hb\ is disfavored as the upper limits to \OIII $\lambda$5007/\Hb\  line ratio  ($<$ 0.08 - 0.5) correspond to metallicities of $\mathrm{12+log[O/H] \geq 8.9}$ \citep{Nagao2006}. Galaxies at $z \sim 1.0$ with such high metallicity should be very massive and bright, and should have been detected in the deep broadband images; 
For LAE-17, 
we lack the spectral coverage to measure \OIII $\lambda$5007 if the detected line were \Hb, but the 
\OII $\lambda$3727/H$\beta$ line ratio ($<$ 0.15)  would again require very low metallicity.

\begin{figure}[tbp]
\begin{center}
\includegraphics[width=90mm]{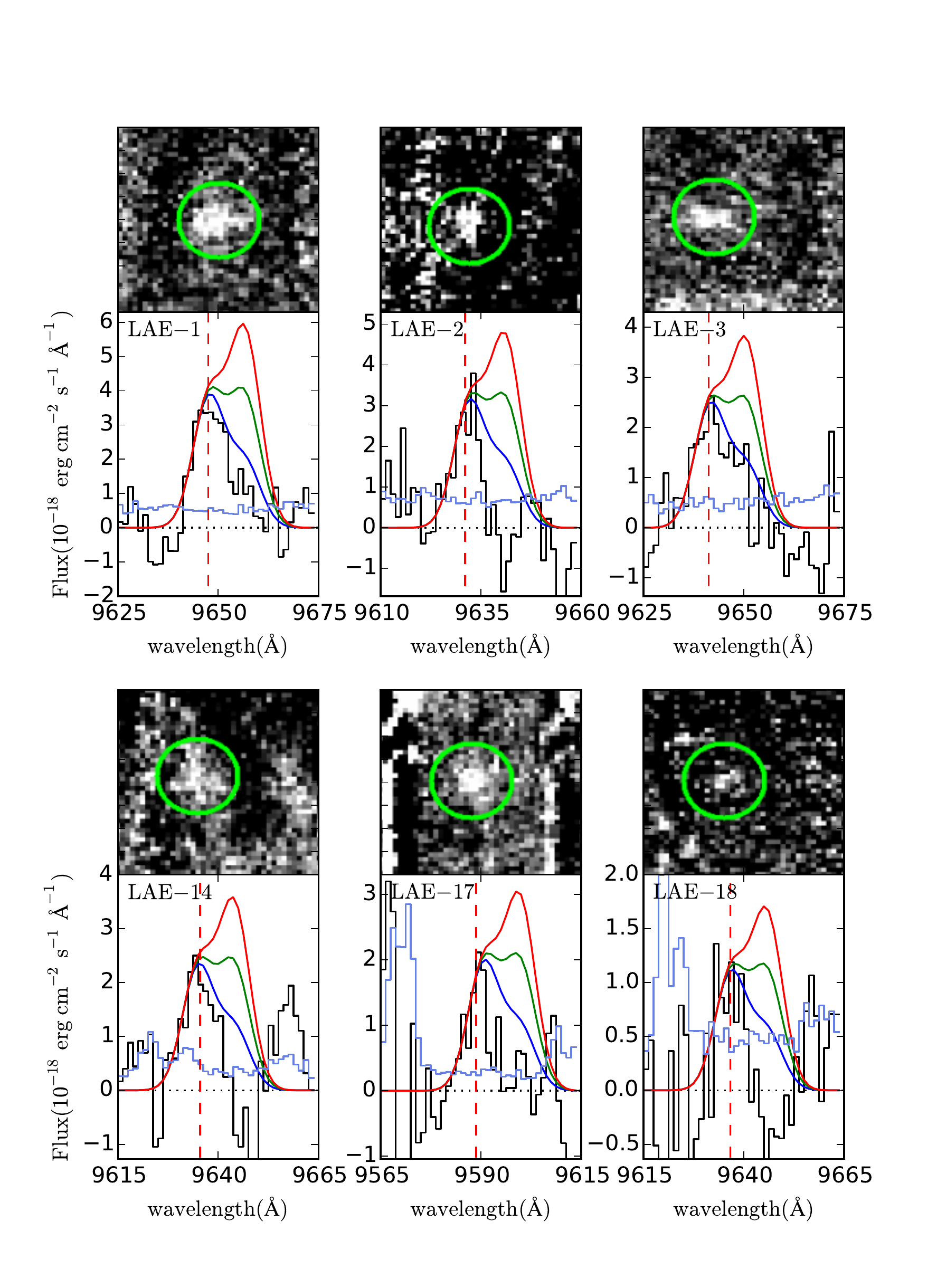}
\caption{\label{OII}Following \citet{Fink2013}, we over plot our line spectra with hypothetical \OII\ doublet lines, assuming three \OII
$\lambda$3726/$\lambda$3729 line ratios of 1.5 (blue), 1.0 (green) and 0.5 (red), respectively. 
{The hypothetical doublet profiles are obtained by fitting the observed line spectra blue ward to the line peaks with  \OII$\lambda$3726.
Obviously most of the lines are too narrow to be \OII\ doublets, meanwhile 
the observed asymmetric line profiles for LAE-1 and LAE-3 are typical for high-$z$ Ly$\alpha$.
Directly fitting the whole range of the line spectra with doublets (allowing the redshift to vary) does not alter the conclusion.}}
\end{center}
%\label{OII}
\end{figure}

If the detected lines are H$\alpha$, then \Hb\ and/or \OIII\ should be detected for each candidate LAE.
Finally, the \OII $\lambda$3727 doublet can also be ruled out.  For LAE-2, LAE-14, LAE-17 and LAE-18, the detected line are too narrow to be \OII $\lambda$3727 (see Fig. \ref{OII}).
If the broader and asymmetric  lines in LAE-1 and LAE-3 were \OII $\lambda$3727, they will have improbably large \OII\ equivalent widths ($>$ 3000 \AA\ in the rest frame, derived with NB964 magnitudes and 3$\sigma$ upper limits to $g$ band magnitude). \OII\ emitters usually have small line EW$_0$s. A sample of 1300 \OII\ emitters at $z \sim 1.5$ have an average line  EW$_0$  of $\sim$ 45 \AA, and the largest line EW$_0$  $<$ 1000\AA\ \citep{Ly2012}.
Furthermore, according to an \OII\ luminosity function \citep{Comparat2015}, we expect $\sim$ 400 \OII\ emitters in our field as bright as LAE-1 and LAE-3.
Therefore, even if \OII\ emitters with EW$_0$ $>$ 3000\AA\  do exist and have a fraction of 1/1300, we only expect 0.3 of them in LAGER COSMOS field.

\begin{figure}[tbp]
\begin{center}
\includegraphics[width=90mm]{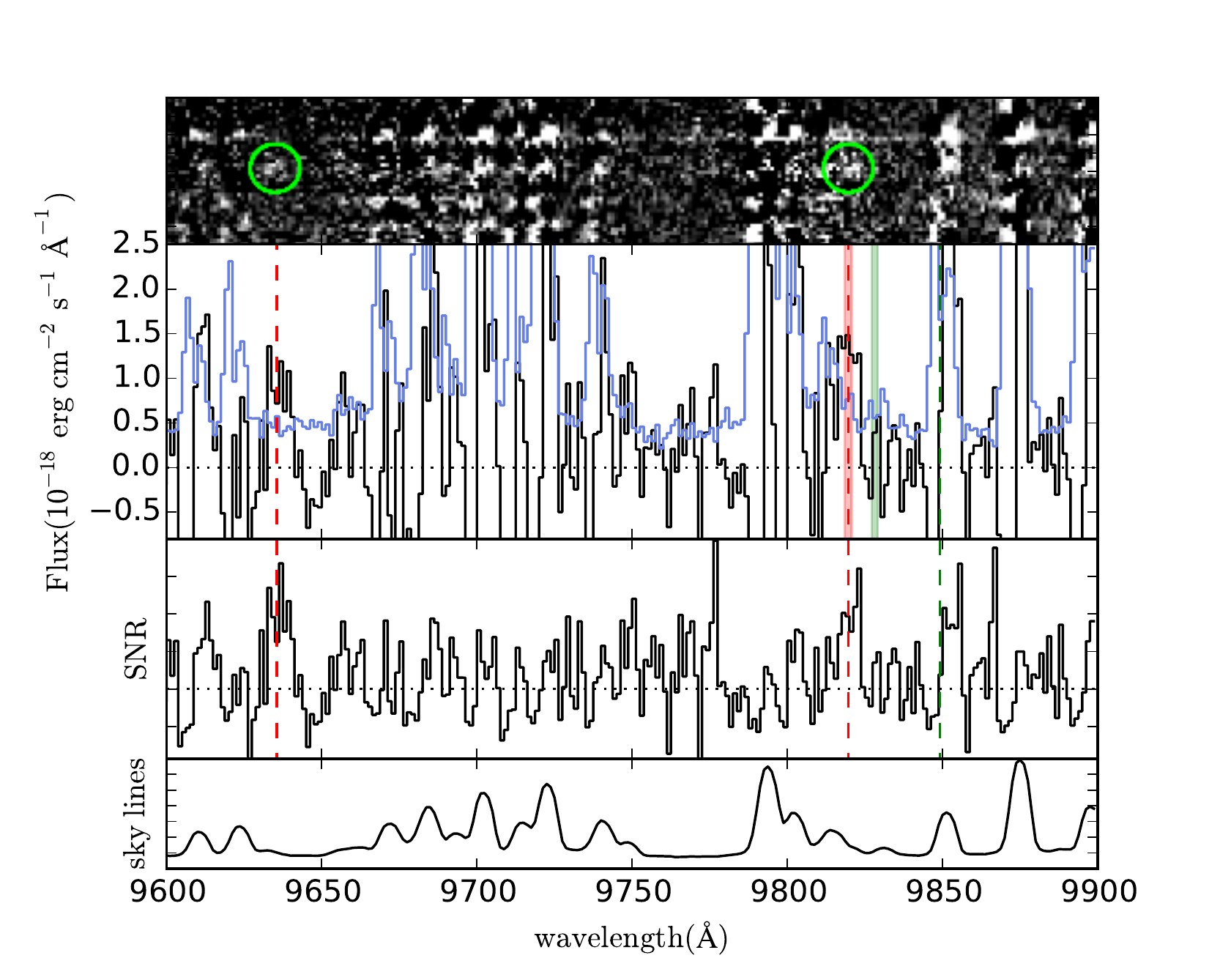}
\caption{\label{NV}A tentative \NV $\lambda$1239 line { at 9819.6$\pm$1.1\AA\ (red shaded region plots the $\pm$ 1$\sigma$ range of the central wavelength)} is seen in LAE-18. The corresponding $\lambda$1243 sits on a sky line (the green vertical line). 
{If the bluer line is \Hb, \OIII$\lambda$4959 is expected at 9828.2$\pm$0.94\AA\ (green shaded region), clearly inconsistent with the observation
(see footnote 2 for details).
}}
\end{center}
%\label{NV}
\end{figure}

In LAE-18, we detect a second emission line at 9819.6$\pm$1.1 \AA\ (see Fig. \ref{NV}). We identify it as \NV $\lambda$1239 line at the same redshift of Ly$\alpha$  ($z = 6.925$)\footnote{ 
{The possibility that the two lines are  \Hb\ ($z = 0.982$) and  \OIII $\lambda$4959 ($z = 0.980$, while \OIII $\lambda$5007 sits on a sky line) respectively can be ruled out, as it would require a blueshift of 
260$\pm$44 km s$^{-1}$ in \OIII\ line. Although large \OIII\ blueshifts are not unusual among AGNs, sources with such large \OIII\ blueshifts always have line width FWHM $>$ 470 km s$^{-1}$ ($\sigma$ $>$ 200 km/s, see Figure 6 in \citet{Bae2014}), much broader than the lines we detected (FWHM $<$ 173 km s$^{-1}$).}},
{with a 3$\sigma$ upper limit of $\sim$ 130 km/s to the potential velocity shift between Ly$\alpha$ and \NV.}
We fit the line with a Gaussian and derive a flux of 1.13$\pm$0.34$\times \mathrm{10^{-17}\ erg\ s^{-1}\ cm^{-2} }$,
{corresponding to a \NV /\lya\ line ratio of 1.33$\pm$0.50. 
Note \citet{Hamann2017} presents an extreme red quasar population which exhibits similarly large \NV /\lya\ ratios at $z\sim2-3.4$.
\citet{Tilvi2016} also shows a tentative \NV\ line with \NV /\lya\ line ratio of 0.85$\pm$0.25 in a $z=7.512$ LBG.}

The Ly$\alpha$ identifications of the six lines are further strengthened by non-detections in the deep broadband images bluer than z band \citep{Zheng2017}. We find no g, r, i detections either using the recently released ultra deep HSC images \citep{Aihara2017}.

We identified neither lines nor continuum from the spectra of the remaining 6 sources. 
For the 3 non-detections on Mask 2 (LAE-11, 21, 23), the sky background is elevated by bright moonlight.  The expected line S/Ns are therefore rather low ($\sim$ 2, 
based on photometric line fluxes, and assuming 50\% slit loss, which is typical for our spectroscopically confirmed LAEs--- cf. table~\ref{tbl:ssp}).  Deeper spectroscopy is therefore still required to clarify whether these sources are in fact  $z \sim 6.9$ LAEs.

For the 3 sources on Mask 1 (LAE-4, 6, 8), we expect to detect Ly$\alpha$ lines with S/N of $\sim$  4 -- 9.\footnote{
{Note the expected line S/N would be considerably lower if the line is intrinsically broader than detected in this work. Assuming a line width of 500 km/s, for LAE-4, 6, 8 the expected line S/N would be $\sim$ 2 -- 4. }
}
Particularly LAE-4 is one luminous LAE candidate, among the 4 we identified.   
These sources are therefore unlikely LAEs, and they could be faint transients instead.

\section{Discussion}

We obtained IMACS spectra for 12  candidate LAEs, and spectroscopically confirmed 6 sources at $z \sim 6.9$, {including 3 luminous LAEs with Ly$\alpha$ luminosities of $\sim$ 10$^{43.5}$ erg s$^{-1}$
and 3 fainter ones at $\sim$ 10$^{42.9}$ erg s$^{-1}$.
The three luminous LAEs have the highest Ly$\alpha$ luminosities among known spectroscopically confirmed  $\gtrsim$ 7.0 galaxies \citep[e.g.][]{Stark2017, Iye2006, Rhoads2012, Ono2012}.}
Considering the expected S/Ns are insufficient for 3 candidates on mask, we achieved a success rate of 67\% (6/9).
Including the $z = 6.944$ LAE J095950.99+021219.1 (which was independently identified both in LAGER and by \citet{Hibon2011}, and spectroscopically confirmed by \citet{Rhoads2012})
we now have a spectroscopic sample of 7 LAEs at $z\sim6.9$ in COSMOS field.
This demonstrates for the first time that narrowband imaging is a highly efficient tool to select LAEs at $z \sim 7.0$, as at lower redshifts.  The high success rate validates the direct use of the photometric sample to derive Ly$\alpha$ luminosity function \citep{Zheng2017}.

\subsection{The Luminous LAEs and Ionized Bubbles}
The spectroscopic confirmations of the 3 (out of 4) luminous LAEs at $z\sim6.9$ are particularly encouraging. 
The bright end excess in the Ly$\alpha$ luminosity function we presented in \citet{Zheng2017} is thus spectroscopically affirmed, supporting a patchy reionization scenario. 
The 4th source showed no signal in IMACS spectra, and is likely a transient. 

The projected distance between the two luminous LAEs (LAE-1 and LAE-3) is rather small (3.6\arcmin, 1.1 pMpc).  Their redshifts differ by only 0.0045, corresponding to a velocity difference of 
170 km s$^{-1}$. Assuming the 3 luminous confirmed LAEs are randomly distributed in the volume our narrowband image probes, the chance probability to have a pair like LAE-1 and LAE-3 is only 0.2\%. 
This indicates LAE-1 and LAE-3 are physically connected.  
They likely sit in a common ionized bubble, which alleviates the neutral IGM attenuation.
Such a bubble could be produced in some combination by the observed LAEs themselves and probably also their close neighbors (such as nearby faint candidate LAE-11) 
(Malhotra et al 2017, in prep). 
These galaxies may have undergone AGN phases (see \S 4.2) which could also have helped produce an ionized bubble. 
The presence of such a bubble could also aid the detection of fainter LAEs, which can be tested with deeper narrowband imaging.

\subsection{The Tentative Detection of N$_V$ Emission Line}

In LAE-18, we detect a possible \NV$\lambda$1240 emission line, with a S/N of 3.3. 
Comparing with \NV, Ly$\alpha$ shows no velocity offset. 

The presence of high ionization \NV$\lambda$1240 requires a very hard ionizing spectrum: 
Production of the \NV\ line  requires ionizing photons of 5.7 Ryd. 
For comparison, the ionization potentials to produce other high ionization UV nebular lines detected in high-$z$ LAEs/LBGs 
are 4 Ryd (\HeII), 3.5 Ryd (\CIV), and 2.5 (\OIII) respectively \citep[e.g.][]{Stark2015b, Stark2015a, Mainali2017, Schmidt2017,Shibuya2017,Sobral2015}.
LAE-18 should thus be a narrow line AGN, driven by a central accreting super-massive black hole.
Furthermore, the presence of \NV\ line also suggests the line emitting gas is metal rich. The line ratios of \NV/\CIV\  and \NV/\HeII\ are often adopted to measure the metallicity in quasars \citep[]{Hamman1993}. Super-solar metallicities have been reported for luminous $z$ $\sim$ 6 quasars \citep{Pentericci2002, Jiang2007, Juarez2009}. It is interesting to find out whether such high metallicity is also seen
in less luminous AGNs at such high redshifts. 
No X-ray or radio counterpart of LAE-18 was detected with COSMOS archival images\footnote{http://irsa.ipac.caltech.edu/data/COSMOS/}, 
{which is perhaps to be expected, as the images are not deep enough to detect a normal AGN at such high redshift}.
Further deep NIR spectroscopic followup is urged to confirm the physical nature of LAE-18 and measure its metallicity. 

{The \NV\ line in LAE-18 can be combined with a possible \NV\ line detected in a $z=7.512$ LBG \citep{Tilvi2016},  and with the detections of high ionization \CIV$\lambda$1549 emission lines (which could also be powered by AGN activity, {or very hot metal-poor stellar population}) in UV selected LBGs at $z \sim$ 6 -- 7 \citep{Stark2015b, Mainali2017, Schmidt2017}.   Note \citet{Shibuya2017} also presented one tentative detection of  \CIV $\lambda$1540 line in one luminous LAE at $z \sim 5.7$.
Meanwhile, the AGN fraction among narrowband selected high redshift (up to $z$ $\sim$ 5.7) LAEs, at Ly$\alpha$ luminosities similar to that of LAE-18,  was known to be very small ($<$ 5 -- 10\%) \citep{Wang2004, Ouchi2008, Wang2009, Zheng2010}.
Together, these results suggest that the AGN fraction among galaxies at $z\gtrsim 6$ may be elevated compared to that at $z\sim$ 4 -- 6. 
If such sources are confirmed as AGNs and are common among $z$ $\gtrsim$ 6 galaxies, they would produce more ionizing photons at a given UV luminosity comparing with typical star forming galaxies, and
could have played a non-negligible role in cosmic reionization \citep[e.g.][]{Giallongo2015,Madau2015}.}

\acknowledgments
{We thank the anonymous referee for useful comments which helped to improve the manuscript.}
We acknowledge financial support from National Science Foundation of China (grants No. 11233002 \& 11421303) 
and National Program for Support of Top-notch Young Professionals for covering the cost of the NB964 narrowband filter. J.X.W. thanks 
support from National Basic Research Program of China (973 program, grant No. 2015CB857005), and CAS Frontier 
Science Key Research Program QYCDJ-SSW-SLH006. Z.Y.Z acknowledges supports by the China-Chile Joint Research 
Fund (CCJRF No. 1503) and the CAS Pioneer Hundred Talents Program (C). 
The work of S.M., J.E.R., and A.G. on this project is supported in part by US National Science Foundation grant AST-1518057.
L.I. is in part supported by CONICYT-Chile grants Basal-CATA PFB-06/2007, 3140542 and Conicyt-PIA-ACT 1417. C.J. acknowledges support by Shanghai Municipal Natural 
Science Foundation (15ZR1446600).

 This research uses data obtained partly through the
Telescope Access Program (TAP), which is funded by the National
Astronomical Observatories, Chinese Academy of Sciences,
and the Special Fund for Astronomy from the Ministry
of Finance. We thank the scientists and telescope operators at
Magellan telescope for their help.

{\it Facilities:}\facility{$Magellan$ (IMACS)}

%\bibliography{ref}

\end{document}